# Sensitivity analysis of an integrated climate-economic model[*]

Benjamin M. Bolker[†], Matheus R. Grasselli[‡], and Emma Holmes[§]

**Abstract.** We conduct a sensitivity analysis of a new type of integrated climate-economic model recently proposed in the literature, where the core economic component is based on the Goodwin-Keen dynamics instead of a neoclassical growth model. Because these models can exhibit much richer behaviour, including multiple equilibria, runaway trajectories and unbounded oscillations, it is crucial to determine how sensitive they are to changes in underlying parameters. We focus on four economic parameters (markup rate, speed of price adjustments, coefficient of money illusion, growth rate of productivity) and two climate parameters (size of upper ocean reservoir, equilibrium climate sensitivity) and show how their relative effects on the outcomes of the model can be quantified by methods that can be applied to an arbitrary number of parameters.

**1. Introduction.** Climate change is recognized as one of the largest risks facing the global financial system. Losses due to extreme weather events alone have risen tenfold in the past 40 years, with a ten-year average now over $200 billion per year. Even the transition to a low-carbon economy poses challenging risks, if only because of the size of the dislocation from carbon-intensive portfolios, with total pledged divestment approaching $15 trillion worldwide. Conversely, the needed investment in green technology, mitigation, and infrastructure is at least an order of magnitude larger than current investment flows, thus presenting a growth opportunity for innovative financial instruments such as green bonds and other green finance initiatives[1]. To adequately address these risks, financial mathematicians need to use and develop models that integrate economic and climate dynamics in a coherent framework. Mainstream models such as the Dynamic Integrated Climate-Economic (DICE) model [12] and its variants make a poor foundation; not only do they generally omit a banking or financial sector, they are also methodologically incompatible with the most salient features of climate science, such as nonlinear feedback mechanisms and tipping points [10]. A better alternative consists of stock-flow consistent models, in which both the economy and the climate are modelled as a system of nonlinear differential equations describing slowly adjusting, out-of-equilibrium quantities [2]. Such models do *not* assume an equilibrium path with a growing economy *a priori*. The dynamics of key economic variables can exhibit runaways, unbounded oscillations and convergence to undesirable equilibria, similar to the behaviour of climate variables in some regimes.

On the other hand, the output dynamics of these models can be very sensitive with respect to parameter values. In particular, due to the possibility of multiple equilibria, small changes in parameters or initial values can lead to completely different long-term values for key economic variables. A preliminary sensitivity analysis of the model in [2] was conducted in [3] in order to investigate the effect of uncertainty in productivity growth, equilibrium temperature sensitivity and a carbon absorption parameter. The purpose of this paper is to extend this analysis in several significant ways.

---

[*]Submitted to the editors DATE.
**Funding:** This work was funded by a Discovery Grant from the Natural Sciences and Engineering Research Council (NSERC) of Canada

[†]Departments of Mathematics & Statistics and Biology, McMaster University (bolker@math.mcmaster.ca)
[‡]Department of Mathematics & Statistics, McMaster University (grasselli@math.mcmaster.ca)
[§]Department of Statistical Sciences, University of Toronto (emmarobin.holmes@mail.utoronto.ca)

[1]Sources: for extreme-weather losses see the Sigma 2 Report by Swiss Re (www.swissre.com). For divestment commitments see Fossil Free (gofossilfree.org/divestment/commitments/). For green investment needs and current flows see the Global Landscape of Climate Finance 2019 Report (www.climatepolicyinitiative.org).

First, we consider the sensitivity of the model with respect to three additional inflation parameters: the speed of price adjustment to fluctuations, the markup rate over labour costs used by firms, and the degree of money illusion in wage bargaining. We establish that the sensitivity with respect to the markup rate is particularly important, because the model converges to two very different equilibria as this parameter varies within the range of empirically observed values.

Based on this observation, we conduct a more detailed analysis to quantify the influence of different parameters on the likelihood of the model converging to an interior equilibrium or one exhibiting explosive behaviour. Specifically, we perform a logistic regression against the several model parameters, with the categorical response variable describing whether the long-term employment rate remains above a given threshold. In this way, we are able to confirm that the markup rate has the largest influence on whether the model converges to an equilibrium with low or high employment in the long run, but this effect decreases when a full feedback from climate change, in terms of both damages and policy responses, is taken into account.

Finally, whereas [3] reports the distribution of some key output variables, such as temperature anomaly and private debt ratio, when parameter values are drawn from their own distributions, it did not describe which parameters contributed the most to the variation in the output. To address this, we use the technique adopted in [1] and compute the partial rank correlation of each parameter under consideration with the employment in year 2100, conditional on it being above a threshold (that is to say, conditional on a 'good' equilibrium). Apart from establishing which parameters are positively or negatively correlated with the employment rate, our results indicate that the effect of uncertainty in economic parameters on the model outputs is comparable with that of uncertainty in climate parameters.

**2. The Model.** We describe the core economic model without climate change first, followed by the full model with climate damages and policy responses.

**2.1. Economic module without climate change.** We adopt the formulation presented in [7], based on the original model proposed in [9], with the necessary modifications to make the model compatible with [2] and [3].

The model makes three key sets of assumptions. The first concerns the relation of output with capital and labour in the economy. We assume that real (i.e inflation adjusted) output is given by $Y = \frac{K}{\bar{\nu}}$, where $\bar{\nu}$ is a constant capital-to-output ratio[2] and $K$ is the real capital stock, which evolves according to $\dot{K} = I - \bar{\delta}K$. Here $I$ denotes real investment by firms and $\bar{\delta}$ is a constant depreciation rate. From real output $Y$, we can obtain the number of employed workers $L = \frac{Y}{a}$, where $a$ denotes the productivity per worker. Denoting the total workforce by $N$, it follows that the employment rate is given by $\lambda = \frac{L}{N} = \frac{Y}{aN}$.

The second set of assumptions has to do with the behaviour of firms. Denote nominal profits by $\Pi = pY - \text{w}L - \bar{r}D$, where $pY$ is total sales revenue of real output $Y$ at a price level $p$, w is the average nominal wage rate per worker, and $\bar{r}$ denotes an average constant rate of interest paid on net debt $D$. The model assumes that real investment by firms is given by $I = \kappa(\pi)Y$ for a function $\kappa(\cdot)$ of the profit share of nominal output

$$\pi = \frac{\Pi}{pY} = 1 - \omega - \bar{r}d, \tag{2.1}$$

where we have introduced the wage share $\omega = \frac{wL}{pY}$ and the debt-to-output ratio $d = \frac{D}{pY}$. In the absence of any other source of financing, firms have to fund this investment either by using profits or borrowing from banks, from which it follows that the change in net debt of firms is given by $\dot{D} = pI - \Pi + \frac{\Delta(\pi)}{pY}$, where the last term denotes dividends paid to shareholders[3].

---

[2]In what follows, constant parameters are denoted with a bar, whereas a dot denotes a time derivative.

The final set of assumptions corresponds to the determination of wages and prices. We assume that the wage rate changes according to $\frac{\dot{w}}{w} = \Phi(\lambda) + \bar{\gamma}i(\omega)$, where $\Phi(\cdot)$, known as the Phillips curve, represents the bargaining power of workers as a function of the employment rate; $\bar{\gamma} \geq 0$ is a coefficient measuring the degree of money illusion (with no illusion corresponding to $\bar{\gamma} = 1$, as is assumed in [2]); and $i(\omega)$ corresponds to the inflation rate, which is assumed to be of the form[4]

$$i(\omega) = \frac{\dot{p}}{p} = \bar{\eta}(\bar{\xi}\omega - 1), \tag{2.2}$$

for an adjustment parameter $\bar{\eta} > 0$ and a markup factor $\bar{\xi} \geq 1$.

Finally, we make two additional assumptions that can be relaxed without altering the model in any significant way, namely that labour productivity $a$ grows exponentially at a constant rate $\bar{\alpha}$ and that the workforce $N$ follows the sigmoid function expressed below in (2.3d). With the assumptions and definitions in place so far, the economy can be described by the following four-dimensional system[5] of coupled nonlinear differential equations for the state variables $(\lambda, \omega, d, N)$:

$$\frac{\dot{\lambda}}{\lambda} = \frac{\kappa(\pi)}{\bar{\nu}} - \bar{\delta} - \bar{\alpha} - \bar{\delta}_N\left(1 - \frac{N}{\bar{N}_{\max}}\right) \tag{2.3a}$$

$$\frac{\dot{\omega}}{\omega} = \Phi(\lambda) - \bar{\alpha} - (1 - \bar{\gamma})i(\omega) \tag{2.3b}$$

$$\frac{\dot{d}}{d} = \frac{\kappa(\pi) - \pi + \Delta(\pi)}{d} - \left[i(\omega) + \frac{\kappa(\pi)}{\bar{\nu}} - \bar{\delta}\right] \tag{2.3c}$$

$$\frac{\dot{N}}{N} = \bar{\delta}_N\left(1 - \frac{N}{\bar{N}_{\max}}\right), \tag{2.3d}$$

where $\pi$ is defined in (2.1) and $\kappa(\cdot)$, $\Phi(\cdot)$ and $\Delta(\cdot)$ are functions that need to be calibrated[6].

---

[3]The original Keen model in [9] does not use dividends, but [2] found necessary to add this term to the model in order to improve the empirical estimates.

[4]The inflation function used in [2] includes the cost of capital and carbon taxes, in addition to labour, as a cost of production for firms. In [3], this is dropped in favour of the simpler inflation dynamics adopted here. Accordingly, the parameters $(\bar{\eta}, \bar{\xi})$ changed from $(0.5, 1.3)$ in [2] to $(0.192, 1.875)$ in [3].

[5]In the original model in [9], the growth rate of $N$ is assumed to be a constant $\bar{\beta}$, so that the relevant dynamics reduces to a three-dimensional system with state variables $(\lambda, \omega, d)$.

[6]For concreteness, we use a linear function for the Phillips curve and truncated linear functions for the investment and dividends, with parameters specified in Table 1.

A full analysis of the equilibria for (2.3) is presented in [7] and summarized here. The interior equilibrium, corresponding to a desirable economic situation of nonzero wages and employment, is given by

$$(2.4) \quad (\lambda^*, \omega^*, d^*, N^*) = \left(\Phi^{-1}\big(\bar{\alpha} + (1-\bar{\gamma})i(\omega^*)\big), 1 - \pi^* - rd^*, \frac{\kappa(\pi^*) - \pi^* + \Delta(\pi^*)}{\bar{\alpha} + i(\omega^*)}, \bar{N}_{\max}\right),$$

where $\pi^* = \kappa^{-1}[\bar{\nu}(\bar{\alpha} + \bar{\delta})]$. As we can see, substituting the expression for $d^*$ into the expression for $\omega^*$ leads to a quadratic equation, from which one can deduce conditions for the existence of at least one strictly positive solution (and sometimes two). There are three other possible equilibria for (2.3), all of which correspond to undesirable economic outcomes. The first corresponds to the state variables $(\lambda, \omega, d)$ converging to $(0, 0, \infty)$, namely vanishing wage share and employment rate and an explosive debt ratio, as first identified in [5]. The second and third represent equilibrium outcomes where the employment rate is zero but the wage share is positive and given by $\omega^{**} = \frac{1}{\xi} + \frac{\Phi(0) - \bar{\alpha}}{\xi \bar{\eta}(1-\bar{\gamma})}$, in which case the equilibrium debt share can be either finite or infinite.

**2.2. Climate Module.** The climate module follows the work of [2], which uses a continuous-time version of the DICE model [12]. It begins by specifying the amount of carbon emissions associated with a level $Y^0$ of industrial production as $E_{ind} = \sigma(1-n)Y^0$, where $\sigma$ is the carbon intensity of the economy and $n$ is an emissions reduction rate. Regardless of the decisions of firms, we assume that technological progress gradually leads the carbon intensity $\sigma$ to decrease in time with a rate $g_\sigma < 0$, which in turn approaches zero at a constant rate $\bar{\delta}_{g_\sigma} < 0$.

To accelerate the transition to an emission-free economy, firms can choose a reduction rate $n$, for which they have to pay abatement costs per unit of production assumed to be of the form $A = \frac{\sigma p_{BS} n^{\bar{\theta}}}{\bar{\theta}}$, where the parameter $\bar{\theta} > 0$ controls the convexity of the cost and $p_{BS}$ is the (inflation adjusted) price of a backstop technology, which we assume to decrease exponentially at a constant rate $\bar{\delta}_{p_{BS}}$. The incentive to pay this abatement cost comes from the fact that firms are assumed to face a carbon tax of the form $T^C = p_C E_{ind}$, where the (inflation adjusted) carbon price $p_C$ increases linearly with a slope $\bar{\delta}_C$. On the other hand, firms also receive a subsidy $S^C = \bar{s}_A A Y^0$ equal to a fraction $0 \leq \bar{s}_A < 1$ of their abatement costs. Accordingly, the emission reduction rate that minimizes the sum of carbon tax and abatement cost is given by

$$(2.5) \quad n = \min\left\{\left(\frac{p_C}{(1-\bar{s}_A)p_{BS}}\right)^{\frac{1}{\bar{\theta}-1}}, 1\right\}.$$

In addition to industrial emissions, the model assumes that there are land-use emissions $E_{land}$ that decrease at a constant rate $\bar{\delta}_{E_{land}} < 0$, so that total emissions are given by $E_T = E_{land} + E_{ind}$. These emissions change the average concentrations of carbon dioxide according to the following three-layer model for the atmosphere, the upper ocean and biosphere, and the lower ocean:

$$(2.6) \quad \begin{pmatrix} \dot{CO_2}^{AT} \\ \dot{CO_2}^{UP} \\ \dot{CO_2}^{LO} \end{pmatrix} = \begin{pmatrix} E_T \\ 0 \\ 0 \end{pmatrix} + \begin{pmatrix} -\bar{\phi}_{12} & \bar{\phi}_{12}\bar{C}_{UP}^{AT} & 0 \\ \bar{\phi}_{12} & -\bar{\phi}_{12}\bar{C}_{UP}^{AT} - \bar{\phi}_{23} & \bar{\phi}_{23}\bar{C}_{LO}^{UP} \\ 0 & \bar{\phi}_{23} & -\bar{\phi}_{23}\bar{C}_{LO}^{UP} \end{pmatrix} \begin{pmatrix} CO_2^{AT} \\ CO_2^{UP} \\ CO_2^{LO} \end{pmatrix}$$

where $\bar{C}_j^i = \frac{\bar{C}^i}{\bar{C}^j}$ for $i,j = \{AT, UP, LO\}$ are ratios of pre-industrial concentrations $\bar{C}^{AT}, \bar{C}^{UP}, \bar{C}^{LO}$ and $\bar{\phi}_{12}, \bar{\phi}_{23}$ are transfer parameters. In turn, an increase in atmospheric $CO_2$ concentration leads to an increase in the Earth radiative forcing, namely the influx of solar energy that is not radiated back into space. Specifically, the model assumes that the increase in radiative forcing $F_{ind}$ caused by industrial emissions is a linear function of the logarithm of the ratio of $CO_2^{AT}$ to its preindustrial level. The planet's total radiative forcing is then given by $F = F_{ind} + F_{exo}$, where $F_{exo}$ is an exogenous increasing function that approaches a constant. Finally, the radiative forcing affects the interplay between the temperature anomaly (compared to pre-industrial levels) $T$ for the atmosphere and upper ocean and the corresponding temperature anomaly $T^{LO}$ for the lower ocean according to the dynamics

$$(2.7a) \qquad \bar{c}\dot{T} = F - \frac{\bar{F}_{dbl}}{\bar{S}}T - \bar{h}(T - T^{LO})$$

$$(2.7b) \qquad \bar{c}^{LO}\dot{T}^{LO} = \bar{h}(T - T^{LO}) \;,$$

where $\bar{F}_{dbl}$ is a parameter that represents the increase in radiative forcing caused by doubling of pre-industrial atmospheric $CO_2$ concentration, $\bar{c}$ and $\bar{c}^{LO}$ are the heat capacities of each layer, and $\bar{h}$ is a parameter representing the heat exchange between layers. It follows from (2.7a)-(2.7b) that the equilibrium climate sensitivity (ECS) – defined as the equilibrium temperature anomaly resulting from a doubling of pre-industrial atmospheric $CO_2$ concentration – is given by the parameter $\bar{S}$.

To summarize, in the absence of any further feedback to the economy, the climate module can be viewed as a component of the model that takes production $Y^0$ as an input and returns the average increase in global temperature $T$ as an output through the following chain of causal relationships explained above: $Y^0 \to E_{ind} \to E_T \to CO_2^{AT} \to F_{ind} \to F \to T$.

**2.3. Coupling the economic and climate modules.** The coupling of the two modules into a unified model is achieved as follows. First, as mentioned in the previous section, for a given level of production $Y^0 = \frac{K}{\bar{\nu}}$, firms incur abatement costs $AY^0$ in order to achieve an emissions reduction rate $n$. These costs are subtracted directly from production, so that only $(1-A)Y^0$ is available as output for sale. Next, because of climate change, a fraction $\mathbf{D}$ of this output is assumed to be irreparably damaged. This fraction is assumed to be a function of temperature anomaly $T$:

$$(2.8) \qquad \mathbf{D} = 1 - \frac{1}{1 + \bar{\zeta}_1 T + \bar{\zeta}_2 T^2 + \bar{\zeta}_3 T^{\bar{\zeta}_4}} \;,$$

where $\bar{\zeta}_1, \bar{\zeta}_2, \bar{\zeta}_3, \bar{\zeta}_4$ are given parameters[7]. Consequently, the output actually sold by firms is given by $Y = (1-\mathbf{D})(1-A)Y^0$. Accordingly, profits for firms need to be modified as

$$(2.9) \qquad \Pi = (1-\mathbf{D})(1-A)Y^0 - wL - \bar{r}D + p(S^C - T^C) \;.$$

where $S^C$ and $T^C$ are the (real) government subsidy and carbon tax mentioned in the previous section, with the correspondingly redefined profit share $\pi = \frac{\Pi}{pY}$.

In order to avoid taking derivatives of the damage function and abatement costs, it is computationally more convenient to use the extensive variables $(K, D, w, p, a, N)$ as state variables for the economic module, instead of the reduced form system in term of the ratios $(\lambda, \omega, d)$, although both these formulations can be shown to be equivalent. Adding the state variables $(\sigma, g_\sigma, E_{land}, CO_2^{AT}, CO_2^{UP}, CO_2^{LO}, T, T^{LO}, p_{BS}, p_C)$ from the climate module described in the previous section leads to a 16-dimensional combined economic-climate model.

---

[7]The desired convexity of such damage curve is contested in the literature: see [4] for critiques of the function used in [12]. In this paper, we assume that the convexity of the damage curve is calculated so that at +4°C anomaly, output is reduced by 10%. This curve is close to that used by [17], and between those used by [12] and [4]. This assumption can be relaxed to allow for varying levels of damage at +4°C.

**Table 1**
*Model parameters*

| Symbol | Value | Parameter description |
|---|---|---|
| $\bar{\alpha}$ | 0.02 | Productivity growth rate |
| $\bar{\delta}$ | 0.04 | Depreciation rate of capital |
| $\bar{\nu}$ | 2.7 | Capital to output ratio |
| $\bar{\delta}_N$ | 0.031 | Work force growth parameter |
| $\bar{N}_{max}$ | 7.065 | Work force equilibrium value |
| $\bar{\Phi}_0$ | -0.292 | Phillips curve y-intercept |
| $\bar{\Phi}_1$ | 0.469 | Phillips curve slope |
| $\bar{\kappa}_0$ | 0.0318 | Investment function y-intercept |
| $\bar{\kappa}_1$ | 0.575 | Investment function slope |
| $\bar{\kappa}_{min}$ | 0 | Investment function minimum |
| $\bar{\kappa}_{max}$ | 0.3 | Investment function maximum |
| $\bar{\Delta}_0$ | -0.078 | Dividend function y-intercept |
| $\bar{\Delta}_1$ | 0.553 | Dividend function slope |
| $\bar{\Delta}_{min}$ | 0 | Dividend function minimum |
| $\bar{\Delta}_{max}$ | 0.3 | Dividend function maximum |
| $\bar{r}$ | 0.02 | Long term interest rate |
| $\bar{\eta}$ | 0.192 | Inflation relaxation parameter |
| $\bar{\xi}$ | 1.875 | Price markup |
| $\bar{\gamma}$ | 0.9 | Effect of inflation on wages |
| $\bar{C}^{AT}$ | 588 | Preindustrial concentration of $CO_2$ in the atmosphere layer |
| $\bar{C}^{UP}$ | 360 | Preindustrial concentration of $CO_2$ in the upper ocean layer |
| $\bar{C}^{LO}$ | 1720 | Preindustrial concentration of $CO_2$ in the lower ocean layer |
| $\bar{\phi}_{12}$ | 0.024 | Transfer coefficient for carbon from AT to UP |
| $\bar{\phi}_{23}$ | 0.001 | Transfer coefficient for carbon from UP to LO |
| $\bar{\delta}_{g_\sigma}$ | -0.001 | Variation rate of the growth of emission intensity |
| $\bar{\delta}_{E_{land}}$ | -0.022 | Growth rate of land use change CO2-equivalent emissions |
| $\bar{\delta}_{p_{BS}}$ | -0.005 | Growth rate of the price of backstop technology |
| $\bar{F}_{dbl}$ | 3.681 | Change in radiative forcing from a doubling of preindustrial $CO_2$ |
| $\bar{F}_{exo}^{start}$ | 0.5 | Initial value of exogenous radiative forcing |
| $\bar{F}_{exo}^{end}$ | 1 | End value of exogenous radiative forcing |
| $\bar{T}_{preind}$ | 13.74 | Preindustrial temperature, in degrees Celsius |
| $\bar{c}$ | 10.20 | Heat capacity of atmosphere and upper ocean layer |
| $\bar{c}^{LO}$ | 3.52 | Heat capacity of the lower ocean layer |
| $\bar{h}$ | 0.0176 | Heat exchange coefficient between temperature layers |
| $\bar{S}$ | 3.1 | Equilibrium climate sensitivity, in degrees Celsius |
| $\bar{\zeta}_1$ | 0 | Damage function parameter |
| $\bar{\zeta}_2$ | 0.00236 | Damage function parameter |
| $\bar{\zeta}_3$ | 4.48e-06 | Damage function parameter |
| $\bar{\zeta}_4$ | 7 | Damage function parameter |
| $\bar{\theta}$ | 2.6 | Abatement cost function parameter |
| $\bar{s}_A$ | 0.5 | Fraction of abatement costs subsidized by government |
| $\bar{\delta}_C$ | 1 | Linear growth rate of the carbon price |

A limited analysis of the equilibria for the full model is provided in [2]. Assuming no inflation, they show that, once the temperature level has reached equilibrium, the 'good' equilibrium – where the economy grows at a constant rate and employment and wages are positive and debt is finite – exists. This equilibrium is similar to (2.4). Further scenario analysis is done numerically in [2] and a sensitivity analysis with respect to the parameters $\bar{\alpha}$ (growth rate of output), $\bar{S}$ (equilibrium climate sensitivity), and $\bar{C}^{UP}$ (size of the intermediate climate reservoir) is provided in [3]. In the next section, we present a more complete sensitivity analysis taking into account some key economic parameters.

**3. Sensitivity Analysis.** We investigate the sensitivity of the economic model without climate change first, followed by an analysis of the full model. Numerical results were obtained by solving the models in R [13] using the package `deSolve` [16] with the `lsoda` integration method[8].

**3.1. Economic Model.** As pointed out in [7], depending on the choice of parameters, the pricing dynamics (2.2) can significantly alter the outcome of the underlying Keen model, because of both the possibility of deflation and the introduction of new undesirable equilibrium points.

To explore the parameter space, the model was run for 180 years, with 20 output time-steps per year, for inflation parameters in the range shown in Figure 1. Points were selected from the parameter space using Sobol sequences implemented by the R package `qrng` [8]. The Sobol method divides the unit interval into successively smaller intervals to generate a quasi-random sequence, which is then generalized to cover a given hypercube with a 'quasi-random' distribution [15].

The outcome of each individual model run was categorized into three possibilities, based on the ending values of the simulation: 'good' if $0.4 \leq \lambda \leq 0.99$, $0.4 \leq \omega \leq 0.99$, and $d \leq 2.7 = \bar{\nu}$, corresponding to an economy with employment and wages bounded away from zero and debt less than the total capital stock[9]; 'outside bounds' if one of $\lambda$ or $\omega$ ends above 0.99; or 'bad' otherwise. The bad outcomes include all of the equilibria corresponding to vanishing employment rate mentioned above, as well as the interior equilibria with low but positive wage shares, which are associated with low inflation (and sometimes even deflation) through (2.2), and consequently low employment rate through (2.4).

The top part of Figure 1 shows the results for the economic module alone, starting with two sets of initial conditions: favourable initial conditions (left), meaning that initially the employment rate and wage share in the economy are high and the debt share is low, and unfavourable initial conditions (right), with lower initial wages and employment and higher initial debt. As we can see in the figure, the key parameter affecting the model outcome is the markup rate $\bar{\xi}$. There are slightly more bad outcomes for low values of the parameter $\bar{\gamma}$ (higher degree of money illusion) or a high relaxation parameter $\bar{\eta}$ (faster price adjustments), but the effects are not pronounced.

An alternative way to explore this result is to look at how changing the markup rate $\bar{\xi}$ changes the basin of attraction to the 'good' equilibrium. Figure 2 shows the results of running the model for a range[10] of initial conditions. The other pricing parameters are fixed at $\bar{\gamma} = 0.9$ and $\bar{\eta} = 0.4$ and we use the same categorized outcomes as in Figure 1. We can see from Figure 2 that increasing the markup rate from $\bar{\xi} = 1.3$ (as used in [2] for an inflation specification that included the cost of capital) to $\bar{\xi} = 1.875$ (as used in [3] for the same inflation specification adopted here) increases the basin of attraction to the good equilibrium (2.4), whereas reducing the markup rate to 1.18 eliminates the possibility of a good economic outcome[11].

For a more detailed analysis, we follow [3] and perform simulations of the model with parameters drawn from appropriately chosen probability distributions[12] fitted to existing empirical estimates.

---

[8]The code used in this paper is available at https://github.com/emmaaholmes/econ-climate-sensitivity and is an extension of the code provided by the authors of [2] and [3]. Any remaining errors are ours.

[9]This also coincides with the 'convergence set' specified in footnote 30 of [2].

[10]Namely, $0.2 \leq \lambda(0) \leq 0.99$, $0.2 \leq \omega(0) \leq 0.99$, $0.1 \leq d(0) \leq 2.7$, which corresponds to the 'initial set' specified in footnote 29 of [2].

[11]If one wants the model to approach the good equilibrium in the case of the good initial conditions and the bad equilibrium in the case of the bad initial conditions, then the markup rate $\bar{\xi}$ should be set around 1.2 – precisely the value used in [7] for illustrations.

[12]For the inflation parameters we use the empirical estimates provided in [6, Online Appendix] and assume that: (1) $\bar{\eta}$ is drawn from a normal distribution with mean 0.4 and standard deviation of 0.12; (2) $\bar{\xi}$ is drawn from a generalized Gamma distribution with shape parameter $s = 3.0894$, scale parameter $m = 0.7154$, and family parameter $f = 0.9959$, shifted right one unit; and (3) $\bar{\gamma}$ is drawn from a generalized Gamma distribution with shape parameter $s = 6.2327$,

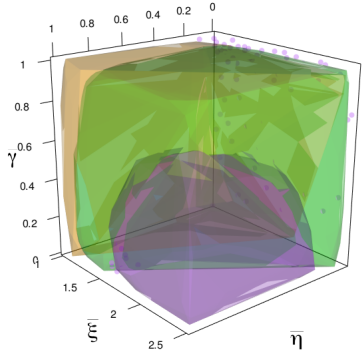
(a) Reduced model, good initial conditions

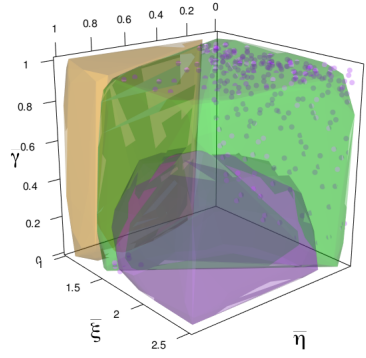
(b) Reduced model, bad initial conditions

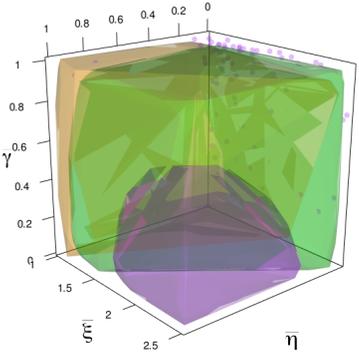
(c) Full model, good initial conditions

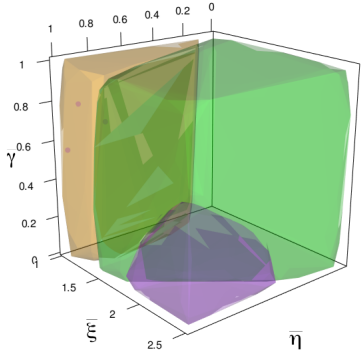
(d) Full model, bad initial conditions

**Figure 1.** *Outcomes of the reduced-form economic model (top panel) and the full model (bottom panel) for a range of inflation parameters and two different sets of initial conditions: $(\lambda(0), \omega(0), d(0)) = (0.9, 0.9, 0.3)$ and $(\lambda(0), \omega(0), d(0)) = (0.5, 0.6, 1)$. A 'good' outcome (green) means the final result satisfies $(\lambda, \omega, d) \in [0.4, 0.99]^2 \times (-\infty, 2.7]$, 'outside bounds' (purple) means that either $\lambda \geq 0.99$ or $\omega \geq 0.99$, and all other outcomes are classified as 'bad' (light orange). Remaining initial conditions match those in [2] and are given in Footnote 16.*

The simulations were conducted by sampling 1000 times from the given probability distributions for the inflation and growth parameters and running the model for each choice using the same initial conditions[13] in 2016. We then sorted the outcome of the model into two categories — good and bad — depending on whether or not the employment rate in 2100 is above 40% and perform a

---

scale parameter $m = 0.0033$, and family parameter $f = 0.3158$, reflected in the y-axis and shifted right one unit. Finally, the productivity growth rate $\bar{\alpha}$ is drawn from a normal distribution with mean 2.06 and standard deviation 1.12 as in [3].

[13]Namely, $(\lambda(0), \omega(0), d(0)) = (0.675, 0.578, 1.53)$, as specified in Appendix D of [2].

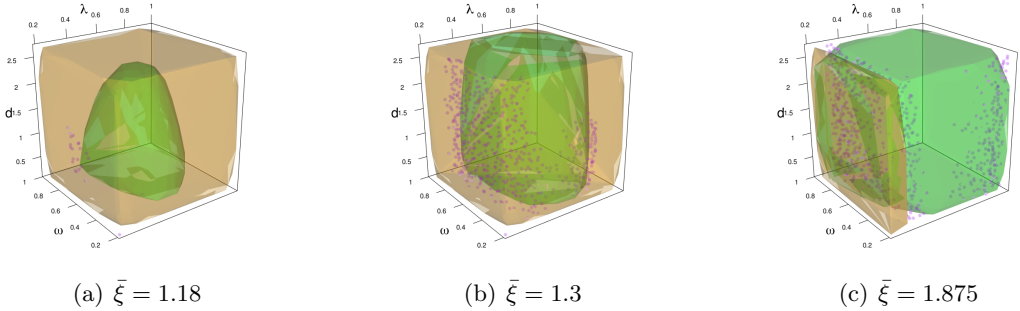

(a) $\bar{\xi} = 1.18$      (b) $\bar{\xi} = 1.3$      (c) $\bar{\xi} = 1.875$

**Figure 2.** *Outcomes of the reduced-form economic model for a variety of initial conditions and three different values of the markup rate $\bar{\xi}$. Increasing the markup rate increases the numerically-computed basin of attraction to the 'good' equilibrium (green), confirming the results shown in Figure 1.*

logistic regression of this categorical outcome with respect to the four parameters[14]. Next we computed the partial rank correlation coefficient[15] (PRCC) associated with each parameter, using the employment rate for good outcomes as the output variable. The results are shown in the top panel of Figure 3 and indicate that the markup rate has a strong effect on the outcome, confirming the conclusions obtained from Figure 1. Moreover, conditioned on converging to a good outcome, the markup rate $\bar{\xi}$ and the money illusion parameter $\bar{\gamma}$ have similar effects on the outcome, but in opposite directions. The relaxation parameter $\bar{\eta}$ and the productivity growth rate $\bar{\alpha}$ have slightly smaller effects.

**3.2. Full Model.** We now repeat the analysis using the full 16-dimensional integrated climate-economic model. As a preliminary result, the bottom part of Figure 1 shows the outcomes of the model using the same sets of initial conditions for $(\lambda, \omega, d)$ and classification criterion as in Figure 1. For the remaining initial conditions we use the values[16] specified in [2]. We observe broadly similar results in Figure 1, indicating that the inflation parameters, in particular the markup factor $\bar{\xi}$, still play an important role in determining the long-term behaviour of the outcomes. Interestingly, comparing the results with unfavourable initial conditions in both cases, we see that the full model exhibits a *larger* range of parameters leading to the good equilibrium, which suggests that, at least for the level of damages[17] and carbon pricing adopted here, the climate module can have a stabilizing effect on the economic module, primarily by preventing excessive investment (through a reduction in profits) and corresponding build-up in private debt.

---

[14]The input variables for the logistic regression were standardized as described in [14] using the R function scale().

[15]PRCCs are used to quantify how the uncertainty in an input variable affects the output, with a higher absolute value indicating a stronger effect [1].

[16] Using $ for 2010 USD and tC for ton of $CO_2$-equivalent, the initial conditions for the full 16-dimensional system are: $K(0) = \$161.3$ trillion, $D(0) = \$91.4$ trillion, $N(0) = 4.83$ billion workers, $w(0) = \$10,591$/(worker-year), $a(0) = \$18,323$/(worker-year), $p(0) = 1$ (normalization constant), $\sigma(0) = 0.6187$ GtC/\$, $g_\sigma(0) = -0.0105$ (year)$^{-1}$, $E_{land}(0) = 2.6$ GtC/year, $p_{BS}(0) = \$547.22$/tC, $p_C(0) = \$2$/tC, $CO_2^{AT}(0) = 851$ GtC, $CO_2^{UP}(0) = 460$ GtC, $CO_2^O(0) = 1740$ GtC, $T(0) = 0.85°$C, $T^O(0) = 0.0068°$C.

[17]Recall that we are using a damage function close to that used by [17] and between those used by [12] and [4].

For a more detailed sensitivity analysis, we again draw model parameters from probability distributions matching available empirical estimates. In addition to the inflation and growth parameters, we follow [3] and consider uncertainty in two climate-related parameters: the equilibrium climate sensitivity $\bar{S}$ and the size $\bar{C}^{UP}$ of the intermediate climate reservoir[18].

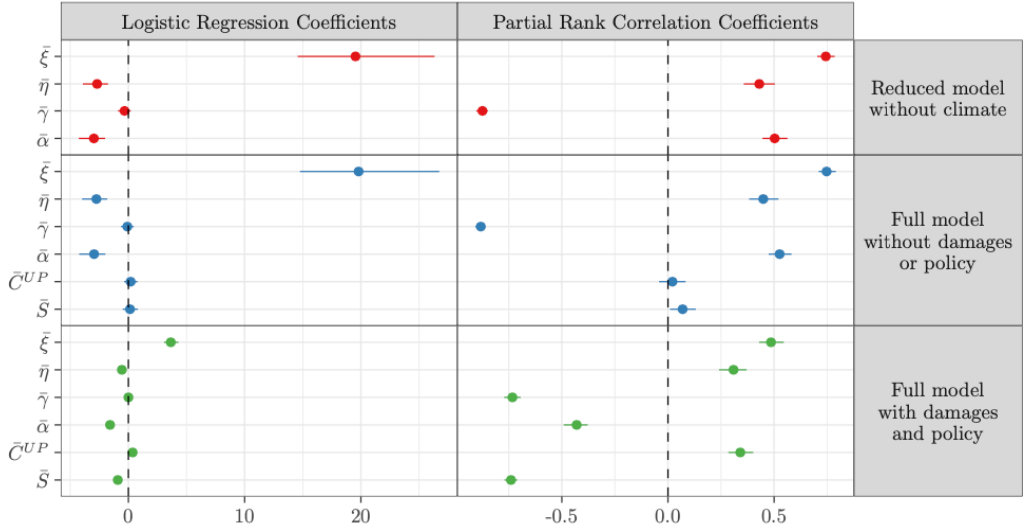

**Figure 3.** *Logistic regression coefficients and partial rank correlation coefficients for the markup rate $\bar{\xi}$, the relaxation parameter $\bar{\eta}$, the money illusion parameter $\bar{\gamma}$, the size of the intermediate climate reservoir $\bar{C}^{UP}$ and the equilibrium climate sensitivity $\bar{S}$.*

As before, the parameters are sampled from their distributions 1000 times and the model is run with the same initial conditions taken from [2]. The same procedure is performed to calculate logistic regression coefficients, where the outcome variable is whether or not the employment rate in 2100 is above 40%, and partial rank correlation coefficients, where the outcome variable is the employment rate in 2100, conditional on being above 40%. The analysis is performed twice, first with the full model but without climate damages and policy, and then again with climate damages, carbon tax and a government subsidy.

The results are shown in the remaining panels of Figure 3. When neither damages nor government policy are taken into account, the logistic regression coefficients again indicate that the markup rate $\bar{\xi}$ has the largest effect, with the labour productivity growth rate $\bar{\alpha}$ and the relaxation parameter $\bar{\eta}$ also having effects significantly different from zero. The PRCC values show that, in this case, uncertainty in the pricing parameters has a greater effect on the outcome than uncertainty in the variables examined in [3], which is unsurprising, given that in this example there is no feedback from the climate module to the economic one.

For the full model with feedback from climate damages, a carbon tax policy and government subsidy, we can see that the climate parameters all have a larger effect, both in determining whether or not the model converges to a good output and in the variability of employment rate in the good outcomes. Interestingly, however, the effects of the inflation parameters are comparable in size to those of the climate ones, meaning that uncertainty in both sets of parameters needs to be taken into account in evaluating the robustness of the model.

---

[18]Specifically, $\bar{S}$ is drawn from a log-normal distribution with mean 1.107 and standard deviation 0.264, and $C^{UP}$ is drawn from a log-normal distribution with mean 5.886 and standard deviation 0.251, which correspond to the distributions used in [3] to match the estimates provided in [11].

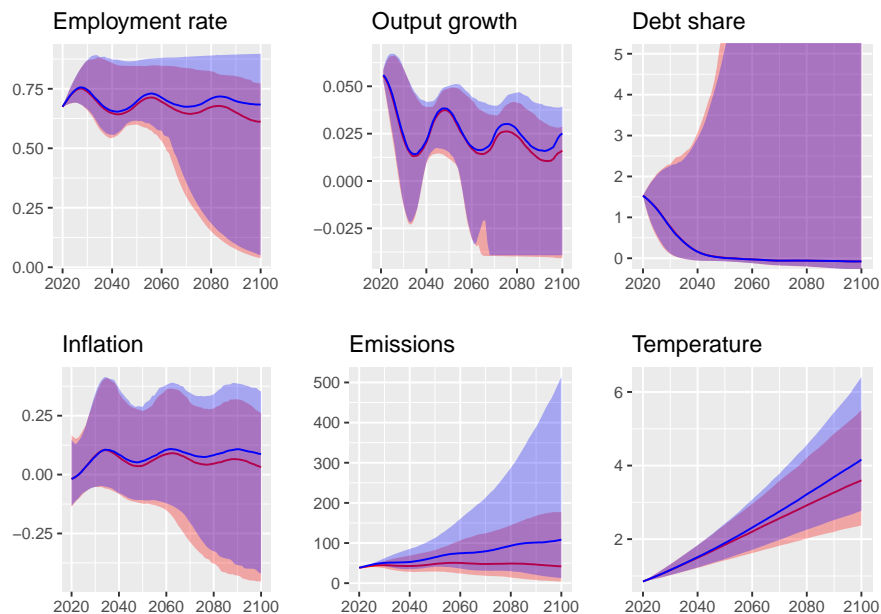

**Figure 4.** *Results of a Monte Carlo simulation of the full model with (in red) and without (in blue) climate damages and government policy. The medians and the 95% confidence intervals are shown for all runs, including both 'good' and 'bad' outcomes.*

As a final illustration, Figure 4 depicts the trajectories of select state variables of the model. Differently from the results presented in Figure 3, this figure shows the values for all runs, instead of only those converging to outcomes with employment higher than 40%, as this allows us to obtain estimates for the unconditional distribution of key state variables. For example, we find that, in the full model with climate feedback and government policy, the temperature anomaly by 2100 has a median value of 3.60 and remains below the Paris accord target of 2°C in 0.8% of the runs, the debt-to-output ratio remains below the acceptable value 2.7 in 86.8% of the runs, and both variables remain below these thresholds in 0.5% of the runs[19].

**4. Conclusion.** In this paper, sensitivity analysis was conducted on a Keen-based model of the climate and the economy. The model is similar to that of [2], but with the pricing dynamics of [7] and slight simplifications to the damage curve and path of the carbon price. The parameter space was explored numerically using Sobol sampling and Monte Carlo methods. We show that the model outcome is sensitive to small changes in some key parameters: the inflation markup rate $\bar{\xi}$, the labour productivity growth rate $\bar{\alpha}$, and the equilibrium climate sensitivity $\bar{S}$.

---

[19]The corresponding values reported in [3] are 2.37°C for the median temperature anomaly, 76.2% of the runs with $d < 2.7$ and 21.5% of the runs with both $T < 2$ and $d < 2.7$. The reason for the difference in results is our use of a more realistic damage curve than the Nordhaus curve adopted in [3], according to which a 4°C temperature increase leads to less than 4% decrease in GDP. The full probability density for the variables shown in Figure 4, similar to Figures 4 and 5 in [3] are easily obtainable from the results of our simulations and available upon request.

These results indicate that sensitivity analyses of integrated climate-economic models need to take into account uncertainties in *all* relevant economic and climate parameters, as they can have significant effects on the conclusions drawn from the model. Accordingly, a future avenue for research would be to estimate these parameter values from existing economic data, as was done in [6], but for this specific version of the Keen model with climate and some of its extensions. Such a study would allow for stronger conclusions about the effectiveness of differing government climate policies, such as carbon price levels and green technology subsidies.

**Acknowledgments.** We are grateful for stimulating discussions with Florent Mc Issac, Etienne Espagne, Devrim Yilmaz, Antoine Godin, Gaël Giraud and the participants of the Georgetown Environmental Justice Program Seminar, where this work was presented.

**REFERENCES**

[1] S. M. BLOWER, D. HARTEL, H. DOWLATABADI, R. M. ANDERSON, AND R. M. MAY, *Drugs, sex and HIV: a mathematical model for New York City*, Philosophical Transactions of the Royal Society of London. Series B: Biological Sciences, 331 (1991), pp. 171–187.

[2] E. BOVARI, G. GIRAUD, AND F. MC ISAAC, *Coping with the collapse: A stock-flow consistent monetary macro-dynamics of global warming*, Ecological Economics, 147 (2018), pp. 383–398.

[3] E. BOVARI, O. LECUYER, AND F. MC ISAAC, *Debt and damages: What are the chances of staying under the 2°C warming threshold?*, International Economics, 155 (2018), pp. 92–108.

[4] S. DIETZ AND N. STERN, *Endogenous growth, convexity of damage and climate risk: how Nordhaus' framework supports deep cuts in carbon emissions*, The Economic Journal, 125 (2015), pp. 574–620.

[5] M. R. GRASSELLI AND B. COSTA LIMA, *An analysis of the Keen model for credit expansion, asset price bubbles and financial fragility*, Mathematics and Financial Economics, 6 (2012), pp. 191–210, https://doi.org/10.1007/s11579-012-0071-8.

[6] M. R. GRASSELLI AND A. MAHESHWARI, *Testing a Goodwin model with general capital accumulation rate*, Metroeconomica, 69 (2018), pp. 619–643.

[7] M. R. GRASSELLI AND A. NGUYEN HUU, *Inflation and speculation in a dynamic macroeconomic model*, Journal of Risk and Financial Management, 8 (2015), pp. 285–310, https://doi.org/10.3390/jrfm8030285.

[8] M. HOFERT AND C. LEMIEUX, *qrng: (Randomized) Quasi-Random Number Generators*, 2019, https://CRAN.R-project.org/package=qrng. R package version 0.0-7.

[9] S. KEEN, *Finance and economic breakdown: Modeling Minsky's "Financial Instability Hypothesis"*, Journal of Post Keynesian Economics, 17 (1995), pp. 607–635, http://www.jstor.org/stable/4538470.

[10] S. KEEN, *The appallingly bad neoclassical economics of climate change*, Globalizations, (2020), pp. 1–29, https://doi.org/10.1080/14747731.2020.1807856.

[11] W. NORDHAUS, *Projections and uncertainties about climate change in an era of minimal climate policies*, American Economic Journal: Economic Policy, 10 (2018), pp. 333–60, https://doi.org/10.1257/pol.20170046.

[12] W. D. NORDHAUS, *Revisiting the social cost of carbon*, Proceedings of the National Academy of Sciences, 114 (2017), pp. 1518–1523, https://doi.org/10.1073/pnas.1609244114.

[13] R CORE TEAM, *R: A Language and Environment for Statistical Computing*, R Foundation for Statistical Computing, Vienna, Austria, 2018, https://www.R-project.org/.

[14] H. SCHIELZETH, *Simple means to improve the interpretability of regression coefficients*, Methods in Ecology and Evolution, 1 (2010), pp. 103–113, https://doi.org/https://doi.org/10.1111/j.2041-210X.2010.00012.x.

[15] I. M. SOBOL', D. ASOTSKY, A. KREININ, AND S. KUCHERENKO, *Construction and comparison of high-dimensional Sobol' generators*, Wilmott, (2011), pp. 64–79.

[16] K. SOETAERT, T. PETZOLDT, AND R. W. SETZER, *Solving differential equations in R: Package deSolve*, Journal of Statistical Software, 33 (2010), pp. 1–25, https://doi.org/10.18637/jss.v033.i09.

[17] M. L. WEITZMAN, *GHG targets as insurance against catastrophic climate damages*, Journal of Public Economic Theory, 14 (2012), pp. 221–244, https://doi.org/10.1111/j.1467-9779.2011.01539.x.